\documentclass[11pt]{article}

\newtheorem{teo}{Theorem}
\newtheorem{pro}{Proposition}
\newtheorem{cor}{Corollary}

\newenvironment{rem}{\par\medskip\noindent\emph{Remark.}}{\par}
\newenvironment{proof}{\par\medskip\noindent\emph{Proof.}
}{\hfill\par}

\newcommand{\cB}{{\cal B}}
\newcommand{\cK}{{\cal K}}
\newcommand{\cR}{{\cal R}}
\newcommand{\C}{{\mathbf C}}
\newcommand{\Z}{{\mathbf Z}}
\newcommand{\ie}{i.\,e.\ }
\newcommand{\ds}{\displaystyle}

\newcommand{\wh}{\widehat}

\newcommand{\p}{\partial}
\newcommand{\intl}{\int\limits}
\newcommand{\suml}{\sum\limits}

\begin{document}

\title{The dynamics of zeros of the finite-gap solutions to the
Schr\"odinger equation}
\author{A. Akhmetshin${}^*$\and Y. Volvovsky${}^{\dagger}$}
\date{}
\maketitle

\begin{center}
{\sffamily Columbia University}\\
{\small
2990 Broadway, Mailcode 4406, New York, NY 10027, USA}\\
and\\
{\sffamily Landau Institute for Theoretical Physics}\\
{\small Kosygina str.~2, Moscow 117334, Russia.}\\
\end{center}

\begin{center}
\small
e-mail:\\
${}^*$\verb"alakhm@math.columbia.edu"\\
${}^{\dagger}$\verb"yurik@math.columbia.edu"
\end{center}

\medskip

\begin{abstract}
Following the recent paper by J.\,F.~van Diejen and H.\,Puschmann
we investigate the behavior of zeros of the finite-gap solutions
to the Schr\"odinger equation. As a result, a new system of
particles on a punctured Riemann surface is constructed. It is
shown to be Hamiltonian and integrable.
\end{abstract}

\section{Introduction}

It was noticed in \cite{dp} that the dynamics of zeros of $n$-solitonic
solutions to the Schr\"odinger equation with the reflectionless potential
is governed by the rational Ruijsenaars\,--\,Schneider system with the
harmonic term (\cite{rs}). This result appears to be surprising since
the aforementioned dynamics was described long ago, though in the
different form. In \cite{dub} it was shown that the solution to
the Schr\"odinger equation
$$
(\p^2_x-u(x))\psi(x,E)=E\psi(x,E)
$$
with the finite-gap potential $u(x)$ is a well-defined function on
the hyperelliptic curve
$$
y^{2}=R_g(E)=\prod_{i=1}^{2g+1} (E-E_i).
$$
The projections $\zeta_j$ of zeros of this function onto the
$E$-plane satisfy the Dubrovin equations (\cite{dub2}):
$$
\frac{\p \zeta_s}{\p x}=\frac{2 \sqrt{R_g(\zeta_s)}}{\prod_{j\ne
s} (\zeta_s-\zeta_j)}.
$$
Notice that these equations contain the parameters of the curve.
The analog of the Dubrovin equations holds also for degenerate
hyperelliptic curves (the latter are described by the same
equation where not all $E_{i}$'s are distinct) and in particular
for fully degenerate hyperelliptic curves which can be thought of
as a Riemann sphere with $n$ couples of pair-wise identified
points. As it was shown in \cite{dp} in the latter case the
parameters of the curve can be excluded from the system. The
modified system then is a system of the second-order differential
equations written solely in terms of zeros of the corresponding
function. It coincides with the Ruijsenaars\,--\,Schneider system
and therefore is Hamiltonian, the expressions for the parameters
of the curve being the integrals of motion.

In this paper we exploit the algebraic-geometrical approach
developed in \cite{kr} to apply these ideas to the case of the
potentials, coming from the hyperelliptic curves with arbitrary
degree of degeneracy.  The dynamics of zeros of the corresponding
solutions to the Schr\"odinger equation is described by the
system, which is shown to be Hamiltonian and completely
integrable, the angle-type variables being the analogs of the
components of the Abel map.

In the second section we study the simplest possible case, when a
hyperelliptic curve is in fact an elliptic curve with several
self-intersections. In the third section we generalize the
results obtained to the case of arbitrary hyperelliptic curve.


\section{Genus one case}

We start with some basic facts from the finite-gap theory.
Consider an elliptic curve $\Gamma$, given by the equation

\begin{equation}\label{cur}
y^2=E^{3}-g_{2}E-g_{3}.
\end{equation}

It's compactified at infinity by one point which we denote by
$\infty$. The only (up to multiplication by constant) holomorphic
differential on $\Gamma$ has the following form: $\ds
\omega^h=\frac{\raisebox{-1pt}{$dE$}}{\raisebox{2pt}{$y$}}$. It
defines the map from $\Gamma$ to the torus
$\wh\Gamma=\C/\Z[2\omega_1,2\omega_2]$, where $2\omega_1$ and
$2\omega_2$ are $a$- and $b$-periods of $\omega^h$, respectively.
This map, given by
$$
A\colon P\longmapsto z=\int_{\infty}^P \omega^h
$$
and known as the Abel map, allows us to identify $\Gamma$ and
$\wh\Gamma$.

Corresponding to the torus $\wh\Gamma$ are the standard
Weierstrass functions
$$\sigma(z|\omega_1,\omega_2),\quad
\zeta(z|\omega_1,\omega_2)=\frac{\sigma'(z|\omega_1,\omega_2)}
{\sigma(z|\omega_1,\omega_2)},\quad \wp(z|\omega_1,\omega_2)=
-\zeta'(z|\omega_1,\omega_2)
$$ (see \cite{bat} for reference).
The function $\sigma(z)$ has the following properties:

i) in the neighborhood of zero $\sigma(z)=z+O(z^{5});$

ii) $\sigma(z+2\omega_j)=e^{2\eta_j(z+\omega_j)}\sigma(z)$, where
$\eta_j=\zeta(\omega_j)$.

Notice that $\wp(z)$ is an elliptic function with the only
(double) pole at $z=0$ and $\zeta(z)$ has the simple pole at
$z=0$ and satisfies the following monodromy conditions:
$$
\zeta(z+2\omega_1)=\zeta(z)+\eta_1,\quad
\zeta(z+2\omega_2)=\zeta(z)+\eta_2.
$$
The map $z\mapsto (E=\wp(z),y=\wp'(z))$ is inverse to the Abel
map.
\medskip

Let us fix $n-1$ points $\kappa_{1},\dots,\kappa_{n-1}$
on~$\wh\Gamma$.

\begin{pro} \emph{\cite{kr1}}\
For generic divisor $D=\gamma_1+\dots+\gamma_n$ on the curve
$\wh\Gamma$ there exists a unique function  $\psi(x, z|D)$
satisfying the following conditions:

1. It's  meromorphic on the curve $\wh\Gamma$ outside the point
$z=0$ and has poles of at most first order at the points
$\gamma_i$, $i=1,\dots,n$.

2. In the neighborhood of $z=0$ it has a form
$$
\psi(x, z)=e^{x z^{-1}}\left(
1+\sum_{s=1}^{\infty}\xi_{s}(x) z^{s}\right).
$$

3. $\psi(x,\kappa_{i})=\psi(x,-\kappa_{i}).$
\end{pro}
\begin{rem}
In general the function $\psi(x,z|D)$ is defined on the curve
$\Gamma$ itself, but here for the sake of brevity we use the
identification between $\Gamma$ and $\wh\Gamma$.
\end{rem}

\begin{proof}
 The uniqueness of such a function follows immediately from
 the Riemann\,--\,Roch Theorem. To show the existence we shall
 consider the following function
\begin{equation}\label{psi}
\psi(x, z|D)=e^{\zeta(z) x}
\frac{\prod_{i=1}^{n}\sigma(z-z_{i}(x))}%
{\prod_{s=1}^n\sigma(z-\gamma_s) \prod_{i=1}^n\sigma(z_i(x))}.
\end{equation}

The set of conditions $\psi(x,\kappa_{i})=\psi(x,-\kappa_{i})$
and the constraint $\sum_{i=1}^{n}z_{i}(x)=x$ (the latter means
that~$\psi$ is an elliptic function) form the system of $n$
equations on the functions $z_{i}(x).$ For generic data this
system is non-degenerate. Then it has the only solution (up to the
permutations) and therefore defines the function $\psi(x, z|D)$
uniquely.
\end{proof}

\begin{cor}
The above-constructed function $\psi(x, z|D)$ is a solution to
the Schr\"odinger equation
\begin{equation}\label{sch}
(\partial_{x}^{2}+u(x))\psi(x, z)=\wp(z)\psi(x, z),
\end{equation}
where $u(x)=-2\sum_{i=1}^{n}\wp(z_{i}(x)) z'_i(x)$.
\end{cor}

\begin{proof}
Consider a function $\psi_0(x,
z)=(\partial_{x}^{2}+u(x)-\wp(z))\psi(x, z)$. It's
straightforward to check that the function $\psi+\psi_0$ satisfy
all defining properties of the function $\psi$. The uniqueness of
$\psi$ implies that $\psi_0=0$.
\end{proof}

\begin{teo}
The zeros of the function $\psi(x,z|D)$ satisfy the following
dynamics:
\begin{equation}\label{soe}
z_{i}''=\sum_{k\ne i}z_{i}'z_{k}'
\frac{\wp'(z_{i})+\wp'(z_{k})}{\wp(z_{i})-\wp(z_{k})}, \quad
i=1,\dots,n .
\end{equation}
\end{teo}

\begin{proof}
To obtain these equations one has to divide (\ref{sch}) by
$\psi(x, z)$ and compare the residues of the both sides of the
obtained equation at the points $z_{i}(x).$
\end{proof}

\begin{rem}
Theorem 1 provides us with a wide class of solutions to system
(\ref{soe}) coming from the algebraic-geometrical data. The simple
"dimensional" argument shows that in fact these are all solutions.
We could reverse the whole reasoning starting with the solution
to (\ref{soe}) and showing that the corresponding elliptic
function (\ref{psi}) solves the Shr\"odinger equation.
\end{rem}

>From now on in this section we shall study system (\ref{soe}). Let
us introduce the variables $\xi_i=\ln z'_i$, $i=1,\dots,n$. In the
variables $z_i$, $\xi_i$ system (\ref{soe}) has the following
form:
\begin{equation}\label{sxi}
\begin{array}{rcl}
z'_i &=& e^{\xi_i},\\
\xi'_i &=&\ds \sum_{k\ne i} e^{\xi_k}
\frac{\wp'(z_{i})+\wp'(z_{k})}{\wp(z_{i})-\wp(z_{k})}, \quad
i=1,\dots,n .
\end{array}
\end{equation}

\begin{pro}
System (\ref{sxi}) is Hamiltonian with respect to the Hamiltonian
$\ds H=\sum_{i=1}^n e^{\xi_j}$ and the 2-form
\begin{equation}\label{o1}
\omega=\sum_{i=1}^n dz_i\wedge d\xi_i-\frac{1}{2}\sum_{i\ne j}
\frac{\wp'(z_{i})+\wp'(z_{j})}{\wp(z_{i})-\wp(z_{j})} dz_i\wedge
dz_j .
\end{equation}
\end{pro}

The proof is a straightforward calculation.

Note that
$$
\begin{array}{rcl}
\omega &=&\ds\sum_{i=1}^n dz_i\wedge d\xi_i-\sum_{j\ne i}
\frac{\wp'(z_{i})}{\wp(z_{i})-\wp(z_{j})} dz_i\wedge dz_j = {}\\
{}&=&\ds\sum_{i=1}^n dz_i\wedge d\xi_i+\sum_{j\ne i} dz_i\wedge
\frac{\wp'(z_j)\,dz_j-\wp'(z_i)\,dz_i}{\wp(z_j)-\wp(z_i)}={}\\
{}&=&\ds\sum_{i=1}^n dz_i\wedge d\xi_i+\sum_{i\ne j}dz_i\wedge
d\Bigl(\ln(\wp(z_j)-\wp(z_i))\Bigr)= \sum_{i=1}^n dz_i\wedge
d\rho_i ,
\end{array}
$$
where
$$
\rho_i=\xi_i+\sum_{j\ne i}\ln(\wp(z_j)-\wp(z_i)) .
$$

The algebraic-geometrical construction from the previous section
provides us with a hint on how the first integrals of system
(\ref{soe}) should look like. The constraints
$\psi(x,\kappa_{s})=\psi(x,-\kappa_{s})$ imply the equations
$$
\sum_{j=1}^{n}\frac{z_{j}'}{\wp(\kappa_{s})-\wp(z_{j})}=0 ,
$$
which can be rewritten in the following form
$$
\sum_{i=1}^n z'_i\prod_{j\ne i} (\wp(z_j)-\wp(\kappa_s))=0 .
$$
These considerations motivate

\begin{teo}\label{T2}
The coefficients $H_k$ of the polynomial
\begin{equation}\label{l}
L(\lambda|z,z')=\sum_{k=0}^{n-1} H_k(z,z')\lambda^k= \sum_{i=1}^n
z'_i \prod_{j\ne i} (\wp(z_j)-\lambda)
\end{equation}
are the integrals of motion of system (\ref{soe}).
\end{teo}

\begin{rem}
Note that the leading coefficient $H_{n-1}(z,z')$ of $L$ is equal
up to the sign to the Hamiltonian $H(z,z')$ of system (\ref{soe}).
\end{rem}

The statement of the theorem is clear since we know that all
solutions are algebraic-geometrical. However, we would like to
present an independent direct proof. It can be found in the
Appendix I.

Let us notice that $L(\wp(z_j))=e^{\rho_j}$. Using this identity
we can rewrite the form $\omega$ in the following way:
$$
\begin{array}{rcl}
\omega &=&\ds\sum_{i=1}^n dz_i\wedge d\rho_i=
\sum_{i=1}^n dz_i\wedge d\ln L(\wp(z_i))={}\\
{}&=&\ds\sum_{i=1}^n \frac{1}{L(\wp(z_i))}\,dz_i\wedge d\left(
\sum_{s=0}^{n-1} H_s \wp^s(z_i) \right)=
\sum_{i=1}^n\sum_{s=0}^{n-1}\frac{\wp^s(z_i)}{L(\wp(z_i))}
\,dz_i\wedge dH_s={}\\
{}&=&\ds \sum_{s=0}^{n-1} d\left(\sum_{i=1}^n \intl^{\wp(z_i)}
\frac{E^s\,dE}{L(E)y(E)} \right)\wedge dH_s+ \sum_{s,k=0}^{n-1}
\left(\sum_{i=1}^n \intl^{\wp(z_i)}
\frac{E^{s+k}\,dE}{L(E)^2y(E)} \right)dH_k\wedge dH_s={}\\
{}&=&\ds \sum_{s=0}^{n-1} d\left(\sum_{i=1}^n \intl^{\wp(z_i)}
\frac{E^s\,dE}{L(E)y(E)} \right)\wedge dH_s ,
\end{array}
$$
where the function $y(E)$ is given by (\ref{cur}).

Thus we have proved the following statement.

\begin{teo}
The variables
$$
\varphi_s=\sum_{i=1}^n \intl^{\wp(z_i)}
\frac{E^s\,dE}{L(E)y(E)},\qquad s=0,\dots,n-1
$$
and $H_s$ defined by (\ref{l}) are the action-angle type
variables for system (\ref{soe}).
\end{teo}

We would like however to rewrite the form $\omega$ once again in
terms of the zeros of the polynomial $L(\lambda|z,z')$ which we
shall denote by~$\wh\kappa_j$, $j=1,\dots,n-1$. In order to do
this we introduce the new variables
$$
\chi_j=\sum_{i=1}^n \intl^{\wp(z_i)}
\frac{dE}{(E-\wh\kappa_j)y(E)},\qquad j=1,\dots,n-1 .
$$
Let us also introduce the variable
$$
\chi=\sum_{i=1}^n \intl^{\wp(z_i)}\frac{dE}{y(E)} .
$$

\begin{teo}\label{T4}
The above-defined form $\omega$ admits the following
representation
\begin{equation}\label{ok}
\omega=d\chi\wedge d(\ln H)+\sum_{j=1}^{n-1} d\chi_j\wedge
d\wh\kappa_j.
\end{equation}
\end{teo}

\begin{rem}
We want to emphasize the fact that the variables
$\{\chi,\chi_j,\,j=1,\dots,n-1\}$ are the degenerate curve
analogs of the components of the Abel map. So our Hamiltonian
structure fits in the general scheme proposed in {\cite{ves}} and
developed in \cite{kp}.
\end{rem}

The proof is a straightforward computation (see Appendix II).


\section{General case}

Consider the hyperelliptic curve $\Gamma$ of genus $g,\,g\ge 1$,
given by the following equation
$$
\Gamma=\Bigl\{Q=(y,E)\ \Bigl|\ y^2=R_g(E)=\prod_{i=1}^{2g+1}(E-E_i)
\Bigr. \Bigr\}.
$$
It's compactified at infinity by one point which we denote by
$\infty$. The curve $\Gamma$ is a $2$-sheeted branched covering
over the complex plane of the variable $E$.

Corresponding to the curve $\Gamma$ is the following system of
differential equations
\begin{equation}\label{niz}
\zeta''_j=\frac{R'_g(\zeta_j)}{2
R_g(\zeta_j)}(\zeta'_j)^2+\sum_{k\ne
j}\frac{\zeta'_j\zeta'_k}{\zeta_j-\zeta_k}\left(1+\sqrt\frac{R_g(\zeta_j)}
{R_g(\zeta_k)}\right).
\end{equation}

\begin{teo}
System (\ref{niz}) is Hamiltonian with respect to the 2-form
$$
\omega=\suml_{j=1}^n \frac{d\zeta_j}{\sqrt{R_g(\zeta_j)}}\wedge
d\rho_j,
$$
where $\ds \rho_j=\zeta'_j+\sum_{k\ne j}\ln(\zeta_k-\zeta_j)$, and
the Hamiltonian $\ds H=\sum_{i=1}^n
\frac{\zeta'_j}{2\sqrt{R_g(\zeta_j)}}$. Coefficients
$H_s(\zeta,\zeta')$, $s=0,\ldots,n-1$ of the polynomial
$$
L(\lambda)=\sum_{k=0}^{n-1}H_s(\zeta,\zeta')\lambda^s=\sum_{j=1}^n
\frac{\zeta'_j}{\sqrt{R_g(\zeta_j)}}\prod_{k\ne
j}(\zeta_k-\lambda)
$$
are the first integrals of system (\ref{niz}). Along with the
functions
$$
\varphi_s=\sum_{j=1}^n \intl^{\zeta_j}
\frac{E^s\,dE}{L(E)y(E)},\qquad s=0,\dots,n-1
$$
they form the set of action-angle type variables for system
(\ref{niz}).
\end{teo}

\begin{rem}
The level sets of the first integrals are not compact. Thus, there is no
canonical choice of action-angle type variables for system (\ref{niz}).
\end{rem}

\begin{proof}
Let a function $\wp_g(x)$ be a solution to the differential
equation
$$
\frac{d\wp_g}{dx}=2\sqrt{R_g(\wp_g)}.
$$
We define new variables $z_j$, $j=1,\ldots,n$ by the conditions
$\zeta_j=\wp_g(z_j)$. In terms of these variables system
(\ref{niz}) has the following form
$$
z_{i}''=\sum_{k\ne i}z_{i}'z_{k}'
\frac{\wp'_g(z_{i})+\wp'_g(z_{k})}{\wp_g(z_{i})-\wp_g(z_{k})},
\qquad i=1,\dots,n .
$$
The rest of the proof is parallel to the genus one case
considered above.
\end{proof}

\begin{rem}
Since $\wp_g(z)$ is a well-defined local parameter on the curve
$\Gamma$ outside infinity, it also follows that in fact
system (\ref{niz}) describes the motion of $n$ particles on
$\Gamma\setminus\infty$.
\end{rem}
\medskip

Our next goal is to show that the dynamics of zeros of solutions
to the Schr\"odinger equation associated with the curve $\Gamma$
is described by the system (\ref{niz}) (the number $n$ of zeros being
bigger than the genus $g$ of $\Gamma$).

The way of constructing these
solutions is standard in the theory of finite-gap operators.

\begin{pro}\label{Pgc}
Let's choose a divisor $R=\kappa_1+\ldots+\kappa_{n-g}$ on the
$E$-plane. For generic divisor $D=\gamma_1+\dots+\gamma_n$ on the
curve $\Gamma$ there exists a unique function  $\psi(x,Q|D,R)$
satisfying the following conditions:

1. It's  meromorphic on the curve $\Gamma$ outside the point
$\infty$ and has poles of at most first order at the points
$\gamma_i$, $i=1,\dots,n$.

2. In the neighborhood of $\infty$ it has a form
$$
\psi(x,Q)=e^{x E^{1/2}} \left(
1+\sum_{s=1}^{\infty}\xi_{s}(x) E^{-s/2} \right).
$$

3. $\psi(x,\kappa^+_{i})=\psi(x,\kappa^-_{i})$, where
$\kappa^{\pm}_i$ are preimages of the point $\kappa_i$ under the
projection onto the $E$-plane.
\end{pro}

\begin{proof}
The uniqueness of such a function (provided it exists) follows
immediately from the Riemann\,--\,Roch Theorem. To show the
existence we write this function down explicitly in terms of the
Riemann $\theta$-function.

The Riemann $\theta$-function, associated with an algebraic
curve~$\Gamma$ of genus~$g$ is an entire function of $g$ complex
variables $z=(z_1,\dots,z_g)$, and is defined by its Fourier
expansion
$$
\theta(z_1,\dots,z_g)=\sum\nolimits_{m\in\Z^g} e^{2\pi i(m,z)+\pi
i (Bm,m)},
$$
where $B=B_{ij}$ is a matrix of $b$-periods,
$B_{ij}=\oint_{b_i}\omega_j$, of normalized holomorphic
differentials~$\omega_j(P)$ on~$\Gamma$: $\oint_{a_j}
\omega_i=\delta_{ij}$. Here $a_i, b_i$ is a basis of cycles on
$\Gamma$ with the canonical matrix of intersections: $a_i\circ
a_j=b_i\circ b_j=0$, $a_i\circ b_j=\delta_{ij}$.

The $\theta$-function has the following monodromy properties with
respect to the lattice $\cB$, spanned by the basis vectors
$e_i\in \C^g$ and the vectors $B_j\in \C^g$ with coordinates
$B_{ij}$:
$$
\theta(z+l)=\theta(z),\qquad
\theta(z+Bl)=\exp\bigl[-i\pi (Bl,l)-2i\pi (l,z) \bigr]\,\theta(z)
$$
where $l$ is an integer vector,  $l\in\Z^g$. The complex torus
$J(\Gamma)=\C^g/\cB$ is called the Jacobian variety of the
algebraic curve~$\Gamma$. The vector $A(Q)$ with coordinates
$$
A_k(Q)=\int_{q_0}^Q \omega_k
$$
defines the so-called Abel transform: $\Gamma\mapsto J(\Gamma)$.

According to the Riemann\,--\,Roch theorem, if the divisors
$D=\gamma_1+\ldots+\gamma_{n}$ and
$\cR=\kappa^+_1+\ldots+\kappa^+_{n-g+1}$, where
$\kappa^+_{n-g+1}=\infty$ , are in the general position then there
exists a unique meromorphic function $r_{\alpha}(Q)$ such that
$D$ is its poles' divisor and
$r_{\alpha}(\kappa^+_{\beta})=\delta_{\alpha\beta}$. It can be
written in the form:
$$
r_{\alpha}(Q)=\frac{f_{\alpha}(Q)}{f_{\alpha}(\kappa^+_{\alpha})},\qquad
f_{\alpha}(Q)=\theta(A(Q)+Z_{\alpha})
\frac{\prod_{\beta\neq \alpha} \theta(A(Q)+F_{\beta})}{%
\prod_{m=1}^{n-g+1}\theta (A(Q)+S_m)},
$$
where
$$
F_{\beta}={}-\cK-A(\kappa^+_{\beta})-\sum_{s=1}^{g-1}
A(\gamma_s),\qquad S_m={}-\cK-A(\gamma_{g-1+m})-\sum_{s=1}^{g-1}
A(\gamma_s),
$$
$$
Z_{\alpha}=Z_0-A(R_{\alpha}), \qquad Z_0={}-\cK-\sum_{s=1}^{n}
A(\gamma_s)+\sum_{\alpha=1}^{n-g+1} A(R_{\alpha}),
$$
where $\cK$ is the vector of Riemann constants.

Let $d\Omega$ be a unique meromorphic differential on $\Gamma$,
which is holomorphic outside $\infty$, where it has double pole,
and is normalized by conditions
$$
\oint_{a_k}d\Omega=0 .
$$
It defines a vector $V$ with the coordinates
$$
V_k=\frac{1}{2\pi i} \oint_{b_k} d\Omega.
$$
Define functions $\psi_{\alpha}(x,Q|D,R)$ by
$$
\psi_{\alpha}= r_{\alpha}(Q)
\frac{\theta(A(Q)+ xV+Z_{\alpha})\;\theta(Z_0)}{%
\theta(A(Q)+Z_{\alpha})\;\theta((xV+Z_0)}
\exp{\left(x\int_{R_{\alpha}}^Qd\Omega\right)}
$$
Let a vector $c_{\alpha}$, $\alpha=1,\ldots,n-g$ be a solution to
the system
$$
\sum_{\alpha=1}^{n-g}c_{\alpha}\psi_{\alpha}(\kappa^-_{\beta})+
\psi_{n+g-1}(\kappa^-_{\beta})=c_{\beta},\qquad
\beta=1,\ldots,n-g.
$$
The function
$$
\psi(x,Q|D,R)=\psi_{n-g+1}(x,Q|D,R)+\sum_{\alpha=1}^{n-g}c_{\alpha}\psi_{\alpha}(x,Q|D,R)
$$
satisfies conditions 1\,--\,3.
\end{proof}

\begin{cor}
The above-constructed function $\psi(x,Q|D,R)$ is a solution to
the Schr\"odinger equation
$$
\partial_{x}^{2}+u(x))\psi(x,Q)=E\psi(x,Q),
$$
where $u(x)=-2\partial_{x} \xi_1(x)$.
\end{cor}

The function $\psi(x,Q)$ has $n$ zeros on the curve $\Gamma$,
whose projections $\zeta_1,\ldots,\zeta_n$ satisfy the set of
Dubrovin equations which in this case read as
$$
\frac{\p\zeta_s}{\p x}=\frac{2
\sqrt{R_g(\zeta_s)}\prod_{\alpha=1}^{n-g}
(\zeta_s-\kappa_{\alpha})}{\prod_{j\ne s} (\zeta_s-\zeta_j)}.
$$
These equations can be rewritten in the matrix form
\begin{equation}\label{mat}
Mv=e,
\end{equation}
where
$$
v_i=\frac{\zeta'_i}{2\sqrt{R_g(\zeta_i)}},\qquad
e_j=\delta_{jg}\qquad\mbox{and}\qquad
M_{ji}=\left\{\begin{array}{ccl}
\ds\zeta_i^{j-1} , &{}&  j\le g ,\\
\ds\frac{1}{\zeta_i-\kappa_{j-g}} , &{}& j>g .
\end{array}\right.
$$

\begin{teo}
Projections $\zeta_j(x)$, $j=1,\ldots,n$ of zeros of the function
$\psi(x,Q|R,D)$, constructed in the Proposition \ref{Pgc}, onto
the $E$-plane satisfy the restriction of system (\ref{niz}) on
the joint level set of its $g$ first integrals:
\begin{equation}\label{res}
H_{n-k}=\delta_{kg},\qquad k=1,\ldots,g.
\end{equation}
\end{teo}

\begin{proof}
Consider a function $\psi(x,Q|R,D)$. Projections $\zeta_j$ of its
zeros onto $E$-plane satisfy matrix equation (\ref{mat}). The
first $g$ equations ensure that restrictions (\ref{res}) are
satisfied. Therefore the polynomial $L(\lambda|\zeta,\zeta')$ is
of degree $n-g$ and the last $n-g$ equations in system (\ref{mat})
state that the points $\kappa_1,\ldots,\kappa_{n-g}$ are its
zeros. The coefficients of $L$ then are time-independent, i.e.
$H'_j(x)=0$, $j=1,\ldots,n$. The last set of equations is
equivalent to system (\ref{niz}).
\end{proof}

\begin{rem}
In fact using the construction from Proposition \ref{Pgc} we can
obtain almost all solutions to the restricted system. Namely, for
general solution $\zeta_j$, $j=1,\ldots,n$ to system (\ref{niz})
satisfying conditions (\ref{res}) we can define
algebraic-geometrical data choosing $R$ to be the zero-divisor of
the polynomial $L$ and $D$ to be
$\ds
\left(\sqrt{R_g(\zeta_1(0))},\zeta_1(0)\right)+\ldots+
\left(\sqrt{R_g(\zeta_n(0))},\zeta_n(0)\right)$.
\end{rem}

\section*{Appendix I}
This appendix contains the proof of Theorem \ref{T2}.
\medskip

Let us consider the polynomial
$$
L(\lambda|z,z')=\sum_{j=1}^n z'_j
\prod_{i\ne j} (\wp(z_i)-\lambda)=
\sum_{k=0}^{n-1} H_k(z,z')\lambda^k .
$$
The explicit formulae for the coefficients $H_k$ are
$$
H_k=\sum_{|J|=n-k-1}(-1)^k \prod_{j\in J}\wp(z_j)\left(
\sum_{k\notin J}z'_k \right) ,
$$
where summation is taken over all subsets
$J\subset\{1,\dots,n\}$ of cardinality $n-k-1$.

We are going to show that the functions $H_k$ are
time-independent, \ie $d H_k/dx=0$. Indeed,
$$
\begin{array}{c}
\ds\frac{d (-1)^k H_k(z,z')}{d x}=
\sum_{J}\prod_{j\in J}\wp(z_i)\sum_{k\notin J} z''_k+
\sum_{J}\left( \sum_{s\in J}\frac{\wp'(z_s)}{\wp(z_s)}z'_s
\right) \prod_{j\in J}\wp(z_i)\sum_{k\notin J} z''_k={}\\
\ds{}=\sum_{J}\prod_{j\in J}\wp(z_j)\left(
\sum_{k\notin J}\sum_{i\ne k}
\frac{\wp'(z_k)+\wp'(z_i)}{\wp(z_k)-\wp(z_i)}\,z'_i z'_k
\right)+
\sum_{J}\prod_{j\in J}\wp(z_j)\left(
\sum_{k\notin J}\sum_{s\in J}
\frac{\wp'(z_s)}{\wp(z_s)}\,z'_k z'_s\right)={}\\
\ds{}=\sum_{J}\prod_{j\in J}\wp(z_j)\left(
\sum_{k\notin J}\sum_{s\in J}\left[
\frac{\wp'(z_s)}{\wp(z_s)}+\frac{\wp'(z_k)+\wp'(z_s)}%
{\wp(z_k)-\wp(z_s)}\right]z'_k z'_s\right)=
\sum_{J,k\notin J, s\in J} \alpha(J,k,s) ,
\end{array}
$$
where
$$
\alpha(J,k,s)=\prod_{j\in J}\wp(z_j)
\left[
\frac{\wp'(z_s)}{\wp(z_s)}+\frac{\wp'(z_k)+\wp'(z_s)}%
{\wp(z_k)-\wp(z_s)}\right]z'_k z'_s  .
$$
Let us consider the involution on the set of triples
$\{J,k\notin J,s\in J\}$ which maps $\{J,k,s\}$ into $\{J',s,k\}$,
where $J'=J\cup\{k\}\setminus\{s\}$.

Now note that
$$
\begin{array}{rcl}
\alpha(J,k,s)+\alpha(J',s,k) &=&\ds
\prod_{j\in J\cap J'}\wp(z_j)z'_k z'_s\Bigl[
\wp'(z_s)+\wp'(z_k)\Bigr.+{}\\
{}&+&\ds\left.
\wp(z_s)\frac{\wp'(z_k)+\wp'(z_s)}{\wp(z_k)-\wp(z_s)}+
\wp(z_k)\frac{\wp'(z_s)+\wp'(z_k)}{\wp(z_s)-\wp(z_k)}
\right]=0
\end{array}
$$
and therefore the whole sum
$\sum_{J,k\notin J, s\in J} \alpha(J,k,s)$ vanishes.

\section*{Appendix II}
Here we present the proof of Theorem \ref{T4}.
\medskip

Consider the 2-form
$\ds\omega=\sum_{s=0}^{n-1} d\varphi_s\wedge dH_s$.
Recall that $H_{s}=(-1)^s H \sigma_{n-s-1}(\wh\kappa)$ for
$s=0,\dots,n-1$, where $\sigma_{n-s-1}(\wh\kappa)$ denotes the
coefficient of $\lambda^s$ in the polynomial
$\prod_{i=1}^{n-1}(\lambda+\wh\kappa_i)$.
By $\sigma_{n-s-2}^j(\wh\kappa)$ we denote the coefficient of
$\lambda^s$ in the polynomial
$\prod_{i\ne j}(\lambda+\wh\kappa_i)$. Then
\begin{equation}\label{ap2}
\begin{array}{rcl}
\omega &=&\ds\sum_{s=0}^{n-1}d\varphi_s\wedge dH_s=
\sum_{s=0}^{n-2} d\varphi_s\wedge
(-1)^s d(H\sigma_{n-s-1}(\wh\kappa))+(-1)^{n-1}\,
d\varphi_{n-1}\wedge dH={}\\
{}&=&\ds \sum_{s=0}^{n-1}(-1)^s \sigma_{n-s-1}(\wh\kappa)\,
d\varphi_s\wedge dH+\sum_{j=1}^{n-1}\sum_{s=0}^{n-2}(-1)^s
H\sigma_{n-s-2}^j(\wh\kappa)\,d\varphi_s\wedge d\wh\kappa_j .
\end{array}
\end{equation}
Now let us notice that
$$
\begin{array}{c}
\ds \sum_{s=0}^{n-2}(-1)^s H\sigma_{n-s-2}^j(\wh\kappa)\,
d\varphi_s= \sum_{s=0}^{n-2}\sum_{l=1}^n(-1)^s H
\sigma_{n-s-2}^j(\wh\kappa)\,d\intl^{\wp(z_l)}
\frac{E^s\,dE}{L(E)y(E)}={}\\
\ds{}=\sum_{l=1}^n d\intl^{\wp(z_l)}
\frac{\suml_{s=0}^{n-2}(-1)^s H \sigma_{n-s-2}^j(\wh\kappa)
E^s}{L(E)y(E)}\,dE -
\sum_{l=1}^n \intl^{\wp(z_l)}d\left(
\frac{\suml_{s=0}^{n-2}(-1)^s H \sigma_{n-s-2}^j(\wh\kappa)
E^s}{L(E)y(E)}\right)\,dE={}\\
\ds{}=\sum_{l=1}^n d\intl^{\wp(z_l)}
\frac{dE}{(E-\wh\kappa_j)y(E)}+
\sum_{l=1}^n \intl^{\wp(z_l)}
\frac{dE}{(E-\wh\kappa_j)^2 y(E)}\,d\wh\kappa_j .
\end{array}
$$
In the same way one can show that
$$
\sum_{s=0}^{n-1}(-1)^s \sigma_{n-s-1}(\wh\kappa)\,d\varphi_s=
\sum_{l=1}^n d\intl^{\wp(z_l)}\frac{dE}{H y(E)}+
\sum_{l=1}^n \intl^{\wp(z_l)}\frac{dE}{H^2 y(E)}\,dH .
$$
Plugging these two formulae into (\ref{ap2}) we obtain
$$
\begin{array}{rcl}
d\omega &=&\ds\sum_{j=1}^{n-1} d\left(\sum_{l=1}^n
\int^{\wp(z_l)}\frac{dE}{(E-\wh\kappa_j) y(E)}\right)
\wedge d\wh\kappa_j+
d\left(\sum_{l=1}^n d\int^{\wp(z_l)}\frac{dE}{H y(E)}\right)
\wedge dH={}\\
{}&=&\ds\sum_{j=1}^{n-1}d\chi_j\wedge d\wh\kappa_j+
d\chi\wedge d(\ln H) .
\end{array}
$$

\section*{Acknowledgements}

The authors are grateful to Professor I.~M.~Krichever for
constant attention to this work.

\end{document}